\documentclass{jaa}
\usepackage[square]{natbib}
\usepackage[colorlinks=true, citecolor = blue]{hyperref}
\usepackage{graphicx}
\usepackage{xcolor}

\begin{document}\sloppy

\title{Detecting cosmological recombination lines with a non-ideal antenna - a first step to practical realization}


\author{Dhashin Krishna\textsuperscript{1,2,*}, Mayuri Sathyanarayana Rao\textsuperscript{1}}
\affilOne{\textsuperscript{1}Astronomy \& Astrophysics, Raman Research Institute, Bengaluru 560080, India.\\}
\affilTwo{\textsuperscript{2}Undergraduate Programme, Indian Institute of Science, Bengaluru 560012, India}


\twocolumn[{

\maketitle

\corres{dhashink@iisc.ac.in}


\begin{abstract}
Photons emitted during the formation of primordial hydrogen and helium atoms over the Epoch of Recombination are expected to be preserved as additive distortions to the Cosmic Microwave Background (CMB) spectrum. The `ripple' like spectral features from Cosmological Recombination Radiation (CRR) have never been detected, and are expected to be 9 orders of magnitude fainter than the CMB. Array of Precision Spectrometers for the Epoch of Recombination - APSERa - is an upcoming ground-based experiment to detect the CRR signal over 2-6 GHz . While astrophysical foregrounds may be theoretically separated from the CRR signal using their inherently different spectral characteristics, instrument generated systematics present a practical problem. We present a first ever study to detect the CRR lines in the presence of a non-ideal antenna adopting a toy model for antenna beam chromaticity. Using Euclidean distance and Pearson correlation coefficient as metrics to distinguish between CRR signal presence and absence in a simulation pipeline, we demonstrate that it is indeed possible to detect the signal using a chromatic antenna. Furthermore, we show that there are different tolerances to the antenna non-ideality based on the type of chromaticity, observing location, and LST. These can inform antenna and experiment design for a practical detection. 
\end{abstract}


}]


\doinum{12.3456/s78910-011-012-3}
\artcitid{\#\#\#\#}
\volnum{000}
\year{0000}
\pgrange{1--}
\setcounter{page}{1}
\lp{1}

\section{Introduction}
Standard cosmology posits that the Universe transitions form being fully ionized to mostly neutral over the cosmological period called the Epoch of Recombination. Doubly ionized helium (HeIII) transitions to singly ionized and eventually neutral helium (HeI) over redshifts spanning $1600 \leq z \leq 8000$ and ionized hydrogen recombines to form neutral atomic hydrogen over $500 \leq z \leq 2000$. The physics of the epoch of recombination has been extensively studied \citep{dubrovich1975,dubrovich2005,kholupenko2006two,Hirata2008,Ali-Haimoud_Hirata_2010} and there exist several codes that simulate the radiative transfer and effective atomic transitions over this period \citep{CosmoSpec,seager2011recfast,HyRec,COSMOREC,Khatri_recomb_code,Rico}. A comprehensive review is in \cite{Chluba}. It is important to note that the redshift of recombination is well after the CMB thermalization redshift of $z\sim10^6$. Consequently any photons that are emitted over recombination are expected to be preserved as an additive distortion to the CMB spectrum. Whereas recombination lines are expected to be transitions resulting in spectral lines, these lines are smoothed due to the extended redshift of recombination as each transitions settles in equilibrium before the next one is effective. Thus the Cosmological Recombination Radiation (CRR) appears as a broad additive spectral distortion to the CMB. Whereas the CRR spectrum is predicted to exist over a wide frequency range, starting as low as $\sim 100$ MHz all the way up to $\sim$ THz, a feasibility study \citep{APSERa2015} (henceforth MSR2015) suggests an octave window between 2-6 GHz is best suited for a detection from the ground. Over this window, the CRR has a characteristic quasi-periodic structure dominated by alpha transitions of the hydrogen atom (redshifted line transitions, n=14-13,13-12,12-11,11-10, and 10-9), which cannot be mimicked by any other astrophysical process. The same feasibility study in MSR2015 introduces Maximally Smooth (MS) functions to describe the foregrounds and provide a method for foreground separation from CRR in a dataset. Thus, MS functions exploit the inherent difference in the spectral properties of the CRR and foregrounds (dominated by Galactic synchrotron emission, and the CMB black-body). On the basis of the feasibility study in MSR2015, an experiment APSERa - Array of Precision Spectrometers for the Epoch of Recombination, has been proposed to operate over a frequency range of $2-4$ GHz. Figure. 1 shows the CRR estimated by CosmoSpec \citep{CosmoSpec}. A detection of the CRR can provide an understanding of the thermal and ionization history of the Universe, and a precise measurement can provide an additional method to determine cosmological parameters \citep{Cosmorec_param_est,param_est_recombination}, have implications for studying annihilating dark matter \citep{Chluba_CRR_DM_2009} and provide an experimental measure of the pre-stellar helium abundance of the Universe \citep{Chluba_2021}. A non-detection would result in a fundamental paradigm shift in cosmology!

\begin{figure}
	\includegraphics[width=\columnwidth]{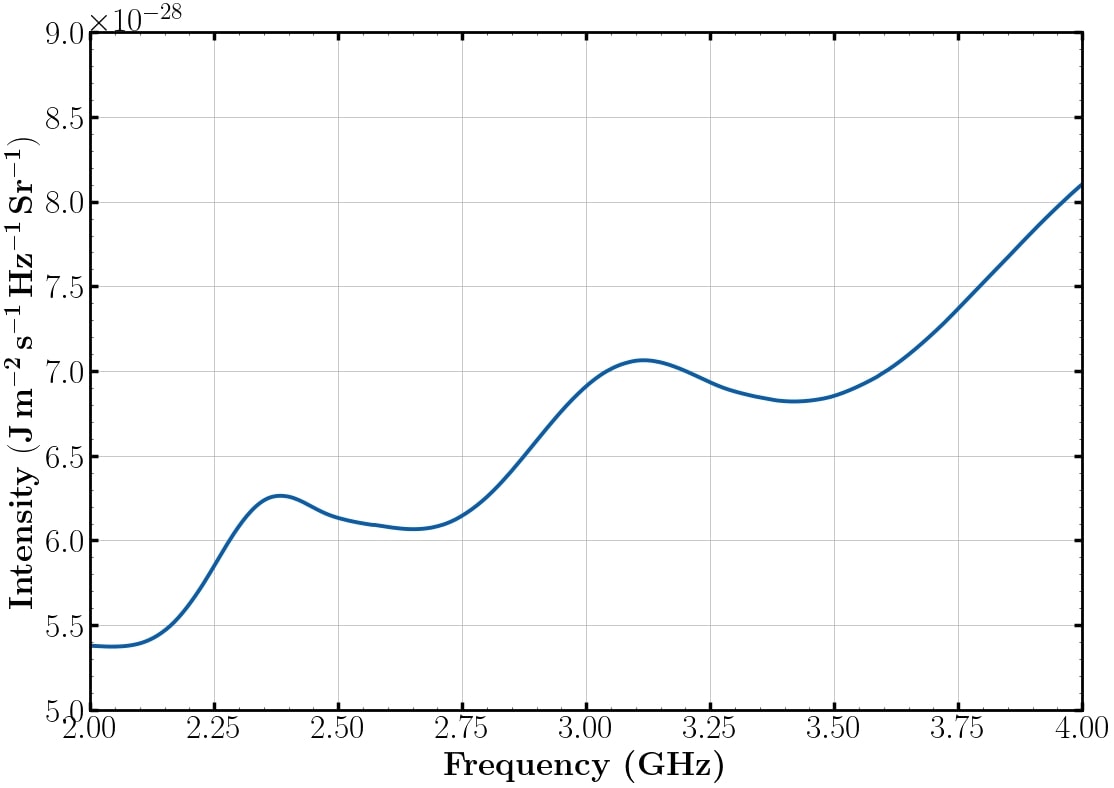}
    \caption{The predicted CRR signal that is present as an additive distortion to the CMB spectrum over 2-4GHz. These ripples are dominated by alpha transitions of primordial hydrogen and a detection can help us study the thermal and ionization history of the Universe. The prediction comes from CosmoSpec \cite{CosmoSpec}.}
    \label{fig:Recombination signal 1.8-4.2GHz}
\end{figure}

Over the 2-4 GHz band, the CRR signal is predicted to have a brightness temperature of $\sim 1-10$~nK. This is $8-9$ orders of magnitude fainter than the CMB. A one part per billion detection is fundamentally challenging. Traditional methods proposed to separate astrophysical foregrounds from global cosmological signals are based on the premise that over sufficiently wide bandwidths foregrounds are spectrally smooth whereas the cosmological signals are not. MSR2015 presented a feasibility study with ideal conditions, including a perfect antenna, and perfectly calibrated receiver, with no instrument induced systematics that can hinder or confuse a signal detection. In practise, any instrument generated systematics that can introduce spurious spectral artefacts that can hinder or confuse CRR signal detection. The most subtle and hence pernicious instrument generated spectral structures are those arising from the chromaticity of the observing antenna. In specific the leakage of spectral structure in the foregrounds into the spectral structure of the measurement set, dubbed ``mode-mixing" \citep{Bowman_2009,Thyagarajan_2016,Morales_2012}, has attracted a lot of attention in the astronomy community, specifically in experiments seeking to detect the redshifted global 21-cm signal from the cosmic dawn and epoch of reionization. Another common contaminant that arises from the antenna are spectrally complex structures due frequency dependent impedance mismatch between the antenna and the receiver. The coupling between the antenna and receiver is parameterized by the antenna return loss, and this too can introduce spectral complexity in the measurement set. Antennas have been designed for the APSERa experiment \cite{discone, Kavitha, sathish2024antenna}. Whereas each antenna has improved over the previous one, it is understandable that achieving achromatic behaviour to one-part-per-billion is challenging if not impossible. \textbf{This paper presents a first ever attempt to simulate antenna-chromaticity using a toy-model beam and provide evidence that within reasonable bounds of specific kinds of chromaticity, the CRR lines can still be detected.} That is, for the first time ever, we demonstrate that though daunting, it may be possible to detect the CRR using an antenna that is not perfectly achromatic, to one-part-per-billion. The simulations demonstrated herein can be used to further inform antenna designs for APSERa, present tolerances in beam chromaticity for a detection and suggest experiment design strategies for a signal detection.

The paper is organized as follows. In Section \ref{sec:pipline} we describe the sky-measurement simulation and signal extraction pipeline. The two metrics of signal detection - namely fractional Euclidean distance and the Pearson Correlation Coefficient, are described in \ref{sec:metric} We describe the beam perturbations in the antenna toy-model and their impact on the signal detection metrics in Section \ref{sec:perturbation}. The results are presented in  \ref{sec:results}, and we close with conclusions and discussions in Section \ref{sec:conclusion}

\section{The pipeline : sky spectrum simulation and signal extraction}\label{sec:pipline}
The pipeline adopted herein broadly follows that described in \cite{sathish2024antenna}, a brief summary is presented here. The antenna-receiver system measures the total intensity from the observable part of the sky at each pixel, weighted by gain of the antenna beam $G$ and multiplied by a windowing function determined by the antenna return loss $\Gamma$. In the case of an antenna with no back lobes (zero response below the horizon, $\theta=0$) the antenna temperature $T_a(\nu,t)$ is presented in equation \ref{Tsky}. 

\begin{equation}
\label{Tsky}
\resizebox{\columnwidth}{!}{$
    {T_a}(\nu ,t) = (1 - {\left| \Gamma  \right|^2})\times \frac{\begin{array}{l}\int\limits_0^{2\pi } {\int\limits_0^{\frac{\pi }{2}} {{T_{sky}}(\theta ,\phi ,\nu ,t)G(\theta ,\phi ,\nu )} } \sin \theta d\theta d\phi \\{\rm{\hspace{0cm}}}\end{array}}{{\int\limits_0^{2\pi } {\int\limits_0^{\frac{\pi }{2}}  {G(\theta ,\phi ,\nu )\sin \theta d\theta d\phi } } }}$}
\end{equation}
\vspace{3mm}

wherein $\Gamma$ is the antenna return loss, $G(\theta ,\phi ,\nu )$ is the antenna beam pattern,  $T_{sky}$ is the sky brightness temperature, $\theta$ is the elevation angle, $\phi$ is the azimuthal angle, $\nu$ is the observing frequency, and $t$ is the observation time.

 In a sky spectrum simulating a null hypothesis, we do not include the CRR, whereas in the case where the CRR signal is present we include the CRR as shown in Figure \ref{fig:Recombination signal 1.8-4.2GHz}. For a diffuse model of Galactic synchrotron emission we adopt a power-law form at each pixel. As inputs to the model we use all sky maps at 408 MHz \citep{haslam408}, 1420 MHz \citep{reich1420_1982,reich1420_1986}and 23 GHz (\textsl{WMAP} science data product\footnote[1]{\textsl{WMAP} Science Team}). The maps are appropriately treated for scaling and offset corrections and are in brightness temperature units. All maps are at Healpix \cite{gorski2005} resolution of $N_{side}=64$ corresponding to a spatial resolution of 0.9161$^{\circ}$. We fit a power law as represented in Equation \ref{eq:power_law}. 
\begin{equation}
    T(\theta_i ,\phi_i, \nu)=T(\theta_i ,\phi_i, \nu_0)  {\left(\frac{\nu}{\nu_0}\right)}^{\beta_i}.
	\label{eq:power_law}
\end{equation}
wherein, $T(\theta_i ,\phi_i, \nu)$ is the brightness temperature of the diffuse emission at pixel $i$ at frequency $\nu$, $\nu_0$ is the reference frequency of 408 MHz, and $\beta_i$ is the spectral index towards pixel $i$. The fitting algorithm aims to minimize the chisquare at each pixel to determine the best fit spectral index. 
We note here that there are several comprehensive and complex sky-model available in the literature \citep{GSM2008,GMOSS2016}. Some of these based on data driven interpolation techniques, such as using Principal Component Analysis, to generate sky-maps at frequencies between the raw input maps. Such models can contain unphysical spectral structure at that can hinder signal detection to 1 part per billion. Adopting more physically motivated models is preferable, However, with intrinsic errors in the input raw maps ranging from $1-10\%$, these errors can propagate through complex equations and further reduce confidence in the overall sky-spectrum. We prioritize signal recovery at a part per billion level from an ideal spectrum (ideal instrument) and demonstrate that even without an ideal antenna metrics of signal detection can be used to recover evience of the presence of the signal. Thus we adopt the power-law model for the sky-model as demonstrated in MSR2015.

\begin{figure}

    \includegraphics[width=\linewidth]{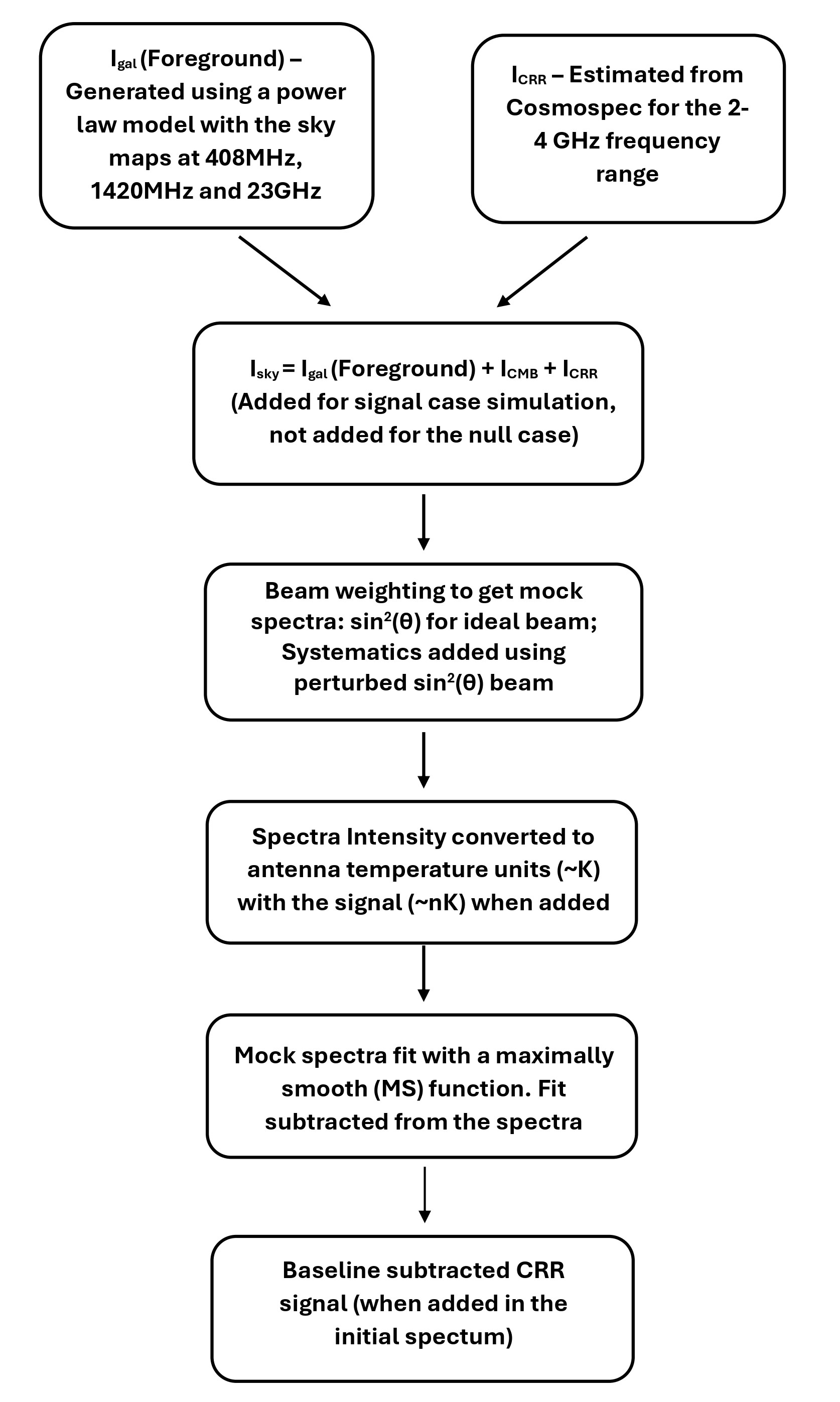}
    \caption{Flowchart describing the APSERa sky simulation and signal extraction pipeline}
    \label{fig:Pipeline flowchart}
\end{figure}

\begin{figure}

    \includegraphics[width=\linewidth]{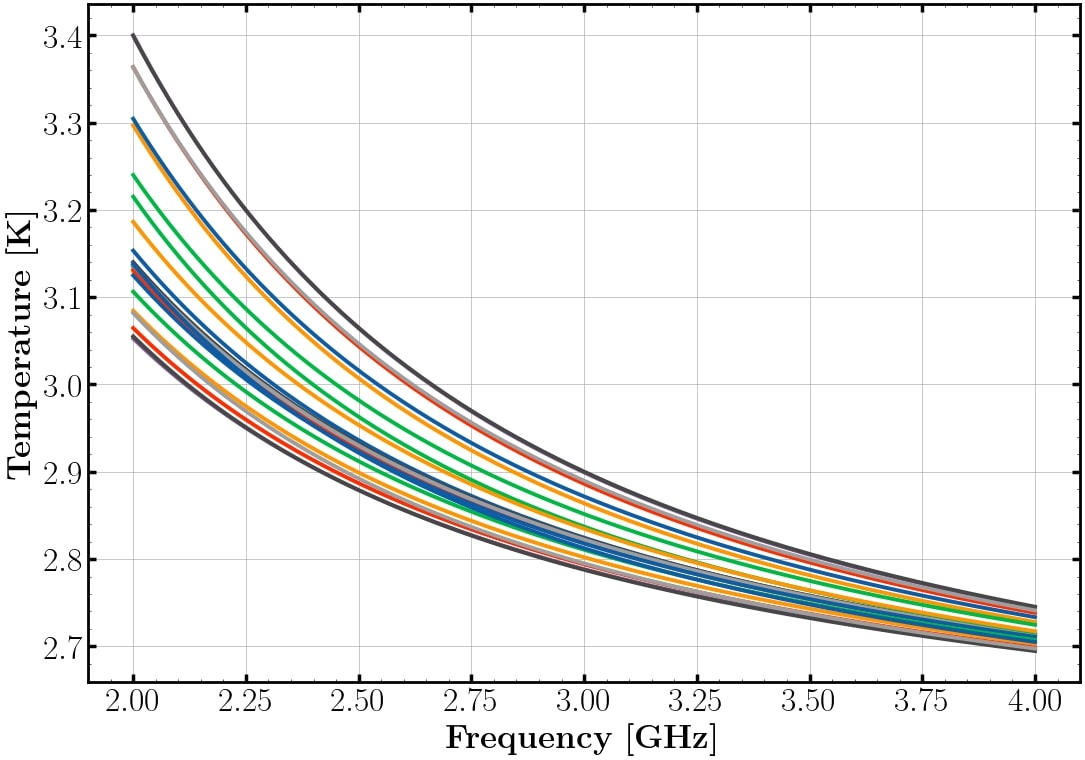}
    \caption{Sample sky spectra generated every one hour for 24 hours in 2-4 GHz frequency range obtained with the ideal $sin^2(\theta)$ beam pattern antenna.}
    \label{fig:24_hours_spectra}
\end{figure}

\begin{figure}[h]
	\includegraphics[width=\linewidth, height = 10cm]{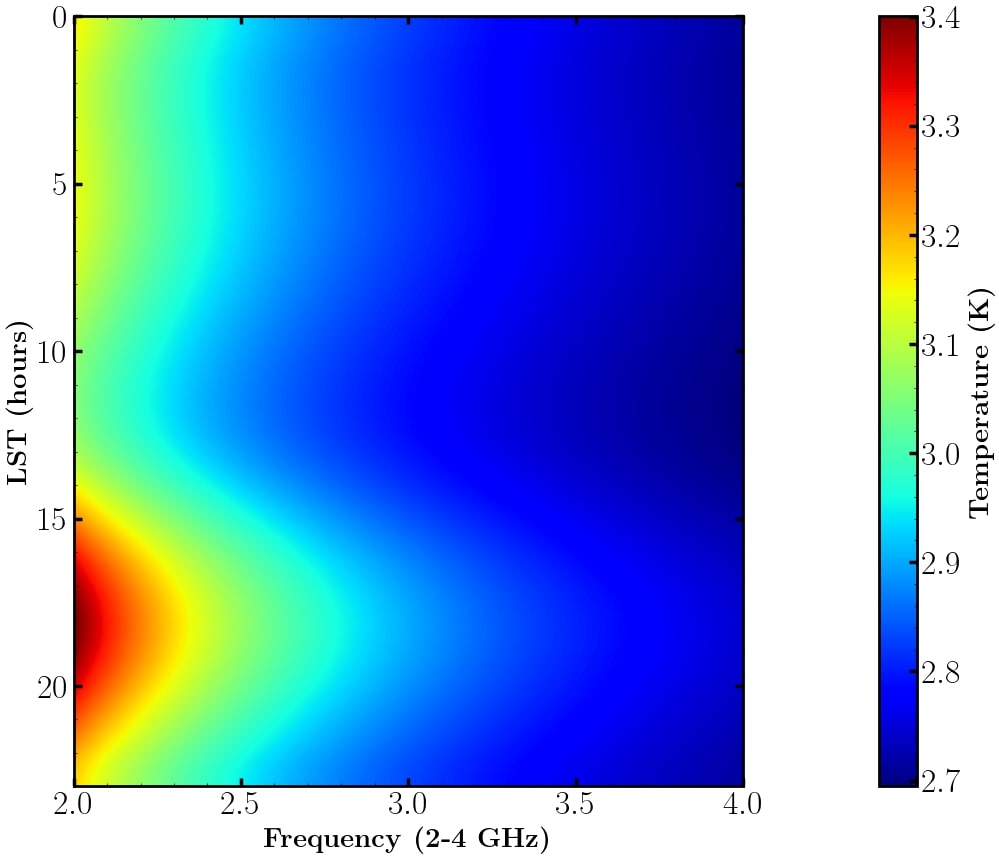}
    \caption{A time-frequency plot showing the variation in the sky spectrum over a period of 24 hours. The waterfall plot clearly illustrates the rising and setting of the Galactic plane.}
    \label{fig:waterfall_plot}
\end{figure}

 A calibration equation is included to convert the beam-weighted sky power to antenna temperature units. The total mock spectrum is of the order of K. It is important to note that in brightness temperature units, the recombination signal is of the order of 10~nK, which is $\sim8$ orders of magnitude less than the foreground brightness temperature. 
We adopt a beam weight of $sin^2(\theta)$, with $0^{\circ}<= \theta <= 90^{\circ}$ being the elevation in degrees above the horizon ($\theta=0^{\circ}$), as the ideal reference antenna pattern. $sin^2(\theta)$ is the analytical beam pattern for a Hertzian or electrically short dipole antenna that exhibits a wide bandwidth and hence a reasonable approximation for the toy model. For the ideal case, the beam pattern remains the same at all frequencies. With this, sample spectra generated at different time stamps over 24 hours and a finely sampled time-frequency water fall plot are shown in Figures \ref{fig:24_hours_spectra} and \ref{fig:waterfall_plot} respectively. The latter captures the effect of rising and setting Galactic centre on the mock spectrum providing a check on the simulation.

The mock spectra so generated are then fit with an MS function of arbitrarily large order (such as order 10). This fitting follows the downhill simplex optimization algorithm to minimize the chisquare of the fit to obtain the MS polynomial coefficients. The order of the MS polynomial is successively increased, starting with a second order polynomial. At each iteration, the optimization solution is accepted (coefficient values) only if the solution respects the criteria that there are no inflection points in the second order derivative of the resulting polynomial. If this criteria for smoothness is violated, then the chisquare is manually reset to a large value to reject the solution and the algorithm continues to search for other solutions. By having a large value (1e5) for the number of iterations that the algorithm will continue to attempt finding a smooth solution to the equation, the code ensures that a large parameter space is explored, overcoming some local minima the final polynomial is smooth. If despite several attempts to find such a solution the algorithm still does not find a smooth solution, then it is a confirmation that the input data contains intrinsic non-smooth features. This is reflected in the large chisquare returned. Any MS function of order 3 or above (order = infinity) will result in the same residual when fitting the CRR signal. Smooth foregrounds on the other hand are completely described to machine precision level by MS functions typically of order 4 to 6 based on LST. MSR2015 has demonstrated that MS functions described the smooth components of a dataset, which in the mock spectra are dominated by foregrounds. The foreground and the baseline smooth component of the CRR are fit by the MS function. The fit is then subtracted to obtain a residual spectrum. This residual spectrum, in the ideal case, contains the baseline subtracted CRR signal in the case when the signal is included in the mock spectrum, and returns the simulated thermal noise in the null hypothesis case.  The baseline subtracted CRR signal is used as a reference template for all further tests - namely the fractional Euclidean distance (denoted by $\gamma$ herein) and the Pearson correlation coefficient ($\varrho$ herein).

\begin{figure}[h]
    \centering
    \includegraphics[width = \linewidth]{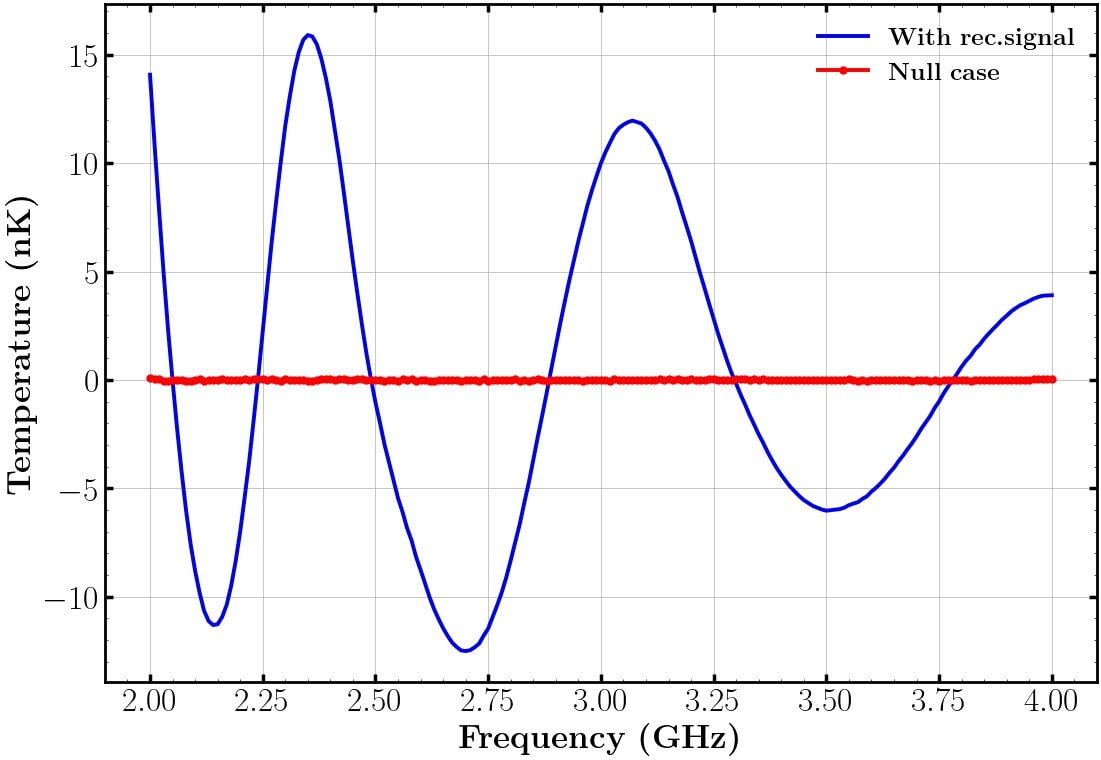}
    \caption{The residuals after fitting an ideal sky spectrum - observed with a perfectly achromatic $sin^2(\theta)$ beam, with the Maximally Smooth (MS) function. The blue residual is for the case when the CRR signal is present in the total sky spectrum, and the red when the signal is absent.}
    \label{fig:Ideal_residual}
\end{figure}

\section{Metrics of signal detection}\label{sec:metric}

The process of the cosmological recombination is well understood and simulations have few (if any) parameters of large uncertainty. Thus there is high confidence in the amplitude and shape of the predicted signal. This serves as the basis of using the template of the expected smooth baseline subtracted signal -- derived from cosmological simulations and passed through the pipeline for MS subtraction --  as a reference for the signal detection metrics. There are no known astrophysical processes in the foreground that can mimic this distinctive signature over the full band. Thus any artefacts that generates similar spectral structure in the total spectrum arise from  instrument non-idealities alone, or pertinent to this work, from the antenna beam chromaticity. We employ two standard measures of similarity as metrics of signal detection, namely Euclidean distance (adapted as a fractional Euclidean distance, elaborated below) and Pearson correlation coefficient. The Pearson correlation coefficient is defined to be scale independent and tracks `similarity' in trends between two variables. The Euclidean distance is applied when the two variables being compared are on the same scale. Whereas the two variables being compared in detection tests below are both residuals on fitting and subtracting mock sky spectra with an MS function, for a test case and reference (ideal) case, we employ both measures of similarity. When the beam perturbation is small, it is expected that the Euclidean distance is a better measure of the similarity between residual in the test and reference cases when the order of magnitude of the residual in both cases are comparable. As the perturbation in the beam is increased the residuals are larger in amplitude compared to the reference case. In such cases, we ask if the mock spectrum continues to retain any indication or memory of the embedded CRR signal and able to discern this from the null hypothesis using the Pearson correlation coefficient metric. Thus, the Euclidean distance may be applied in cases when the residuals in the test and reference case are comparable in amplitude (likely when the beam perturbation is low) and the Pearson Correlation coefficient may be applied to test if the signal is present or absent when the residual in the test case(s) are no longer comparable in amplitude to the reference case.

\subsection{Euclidean distance}
The Euclidean Distance between two points, a standard measure in mathematics, is the length of the connecting line segment between the two points in Euclidean space. We adopt this formalism to describe the Euclidean distance between two signals, or effectively two arrays of data points, namely the residual obtained in the test case and the reference case. The test case signal is the residual obtained by subtracting the mock spectrum (with or without the CRR signal) by a best fit MS function wherein the mock spectrum is itself generated using a non-ideal or `perturbed' beam. The reference case is always the ideal CRR signal residual obtained using a perfectly achromatic $sin^2(\theta)$ beam, shown in Figure \ref{fig:Ideal_residual} as the blue solid line.

 It is more meaningful to look at the Euclidean distance not as an absolute number, but as a relative number when applied to the test cases that follow. A meaningful detection of the signal can only be claimed if there is a clear distinction between the presence and absence of the signal. If beam induced systematics creates artefacts in a null hypothesis spectrum suggesting a signal detection, this results in a false positive.  Thus, we define the fractional Euclidean distance $\gamma$ as follows.
 
\begin{equation}
    \gamma = \dfrac{ED_{null} - ED_{sig}}{ED_{null}}.
\end{equation}\label{eq:gamma}

wherein, ED$_{sig}$ and ED$_{null}$ are the Euclidean distance of the test cases from the reference signal, residual with the CRR present and absent (null hypothesis) respectively. We note that this definition of $\gamma$ is conservatively defined towards rejecting a false positive, but not towards detecting or rejecting a false negative. 

In this definition, the closer the fractional Euclidean distance is to 1, the stronger the indication is for presence of the signal than absence (marginalizing over artefacts of beam induced systematics suggesting a false positive). 

In the extreme case when the perturbation to the beam is zero, i.e. the test case is the same as the reference case, we expect the Euclidean distance for the signal present to be zero and that for the null case to be large, resulting in $\gamma=1$. Whereas $\infty<\gamma<1$, in the limiting case where no distinction can be made between signal presence and absence $\gamma=0$. That is, when the perturbation in the beam is very large, and thus there is significant spectral feature in the mock spectrum, the residual in both cases (CRR present and absent) will look identical due to the limiting nature of MS functions - the fitting algorithm saturates when it hits the limit of the first sign of non-smoothness perhaps orders of magnitude above the CRR signal amplitude.

\subsection{Pearson's correlation coefficient}
Pearson's correlation coefficient ($\varrho$) is a measure of linear correlation between two variables. In this case the two variables are the residuals from the test case $x$ and the reference case $y$, with means $\Bar{x}$ and $\Bar{y}$ respectively. $\varrho$ is defined in Equation \ref{eq:PCC}.
\begin{equation}\label{eq:PCC}
    \varrho = \dfrac{\sum_{i=1}^n (x_i - \Bar{x})(y_i - \Bar{y})}{\sqrt{\sum_{i=1}^n(x_i - \Bar{x})^2}\sqrt{\sum_{i=1}^n(y_i - \Bar{y})^2}}
\end{equation}

 $\varrho$ ranges from -1 to 1, namely from perfectly anti-correlated, 0 for uncorrelated and to 1 for perfectly correlated.  Given the well defined range of $\varrho$, we can separately investigate the value $\varrho$ takes in the tests cases when the signal is present and when the signal is absent. In the extreme cases, when there is no perturbation and the beam in the test case is identical to the reference case, $\varrho=1$ for the case when the signal is present, and $\varrho=0$ when the signal is absent. When the perturbation is very large $\varrho$ can range from $-1\leq \varrho \leq1$ in both cases -  resulting in an inconclusive detection suggesting false negatives when the signal is present and a false positive when the signal absent. 

Thus the fractional Euclidean distance and Pearson's correlation coefficient provide useful complementary metrics to signal detection, depending on the amount of perturbation or non-ideality in the beam, as demonstrated in Section \ref{sec:results} We now discuss the different beam perturbations built in the toy model to investigate the effect of antenna chromaticity on the above defined signal detection metrics.

\section{Beam perturbations and their effects} \label{sec:perturbation}
Any antenna in the real world exhibits varying levels of chromaticity, such as a beam that varies with frequency, especially when operating over a broad range of frequencies such as one full octave. Whereas there are multiple ways in which the beam may change with frequencies that are typically complex, we consider two specific forms of beam variation as a toy model and ask the question, is it possible to detect the CRR lines with such `non-ideal' antenna behaviour. Specifically we term the two beam frequency dependent perturbations as (i) wobbling with two sub-cases and (ii) stretching. While simplistic, this exercise is a first step in moving away from the ideal beam used in the feasibility study in MSR2015.

As mentioned previously, the analysis here considers the $sin^2(\theta)$ beam pattern over the full band of 2-4 GHz as the ideal case, where $\theta$ is the elevation angle defined as zero at the horizon. The scenarios considered to introduce beam chromaticity in the toy model are (i) changing the half power beam width at different frequencies - called beam stretching henceforth and (ii) tilting of the beam maximum direction with frequency - called beam wobbling. The beam wobbling can be realized as two distinct sub-cases, where the beam strictly wobbles in one direction with frequency - called 1-D(irectional) beam wobbling henceforth and tilting of the beam maximum back and forth - called 2-D(irectional) beam wobbling henceforth. In both 1-D and 2-D wobbling, the central frequency of 3 GHz maintains a $sin^2(\theta)$ beam pattern with the beam peaking towards the zenith. A visual representation of the beam perturbation for each case is available in the links in the respective subsections below.

\subsection{Beam `wobbling'}
The reference antenna beam of $sin^2(\theta)$ peaks at an elevation angle $\theta=90^{\circ}$. This is chosen as the beam at the mid-frequency of 3 GHz. Varying degrees of beam wobble are introduced to change the direction of the beam maximum for different levels of perturbation. The `amount' of beam wobble is parameterized by $\alpha$. Beam wobbling introduces contributions from foregrounds at different directions in the sky (effectively different foreground pixels) at different frequencies with varying beam weight, generating a mock spectrum where spatial features produce spectral features. The Galactic emission is itself not very bright over $2-4$ GHz, as is the case at much lower frequencies ($\sim 100 MHz$), the effect of beam wobble is not very obvious in the total spectrum. The effect of the chromaticity induced spectral structure is seen instead in the residuals of the test cases.

For a wobble of $\alpha^{\circ}$, we model the beam using the transformed coordinate corresponding to a rotation of $\alpha^{\circ}$. The modified beam weight in the toy model is realized using equations \ref{eq:wobble1},\ref{eq:wobble2} when the beam wobbles. 
\begin{equation}\label{eq:wobble1}
G(\theta, \phi, \nu) = \mathrm{sin}^2(\theta'),  \text{where}
\end{equation}  

\vspace{-6mm}
\begin{equation}\label{eq:wobble2}
    \theta' = \mathrm{cos}^{-1}(\mathrm{sin}(\theta) \mathrm{sin}(\phi) \mathrm{sin}(\alpha_{\nu}) + \mathrm{cos}(\theta) \mathrm{cos}(\alpha_{\nu}))
\end{equation}
\vspace{-6mm}
\begin{equation}\label{eq:wobble3}
    \phi' = \mathrm{tan}^{-1}(\mathrm{tan}(\phi) \mathrm{cos}(\alpha_{\nu}) - \mathrm{cot}(\theta) \mathrm{sin}(\alpha_{\nu}) \mathrm{sin}(\phi))
\end{equation}

To introduce frequency dependence in the 1-D beam wobble case, the total angle of the wobble $\alpha$ take the form in equation \ref{eq:wobble4}
\begin{equation}\label{eq:wobble4}
    \text{Wobble angle }(\nu_{GHz}) = \alpha_{\nu} = \alpha \left(\dfrac{\nu_{GHz} - 3}{2}\right).
\end{equation}

$\alpha$  parameterizes the total range in the peak elevation angle over the frequency range of 2-4 GHz, with the shift being equally distributed between $-\alpha/2$ to $\alpha/2$ around the zenith ($\theta=90^{\circ}$).

In the case of 2-D wobble, the frequency dependence takes a modified form given by  \ref{eq:wobble5}
\begin{equation}\label{eq:wobble5}
    \text{Wobble angle }(\nu_{GHz}) = \alpha_{\nu} = \alpha \left(\dfrac{|\nu_{GHz} - 3|}{2}\right).
\end{equation}

\begin{figure}[h]
    \centering
    \includegraphics[width = 0.9\linewidth]{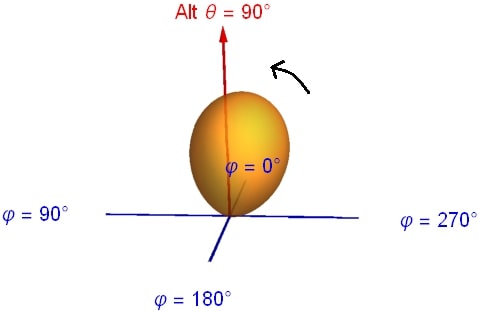}
    \includegraphics[width = 0.9\linewidth]{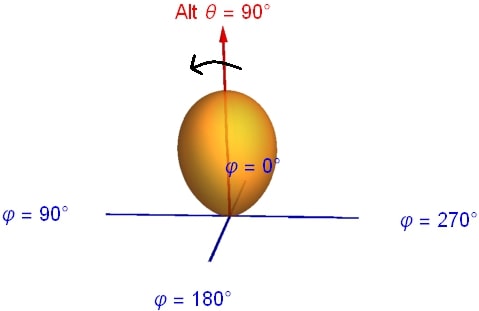}
    \includegraphics[width = 0.9\linewidth]{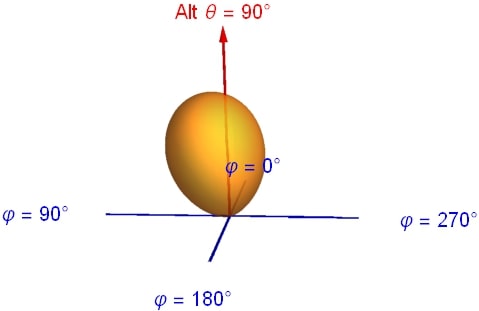}
    \caption{The perturbed beam showing 1-directional wobble at spot frequencies of 2GHz, 3GHz, and 4GHz, with a total wobble angle of 20$^{\circ}$. }
    \label{fig:Wobble_beam}
\end{figure}

\begin{figure}[h]
    \centering
    \includegraphics[width = \linewidth]{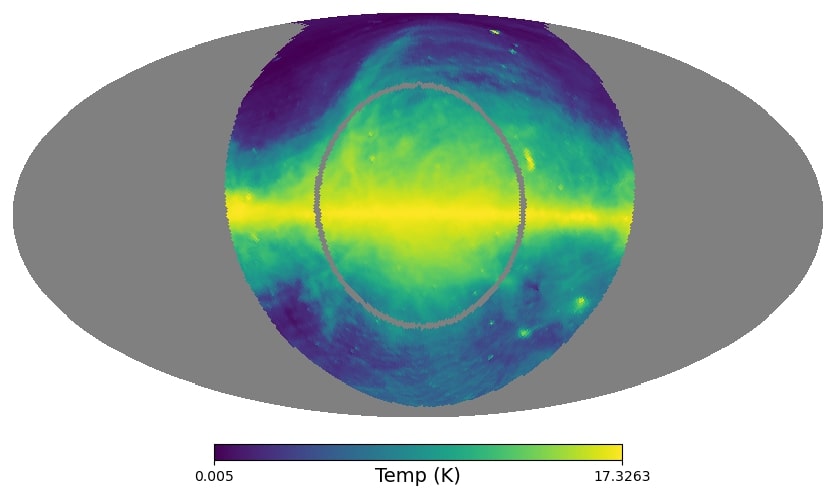}
    \includegraphics[width = \linewidth]{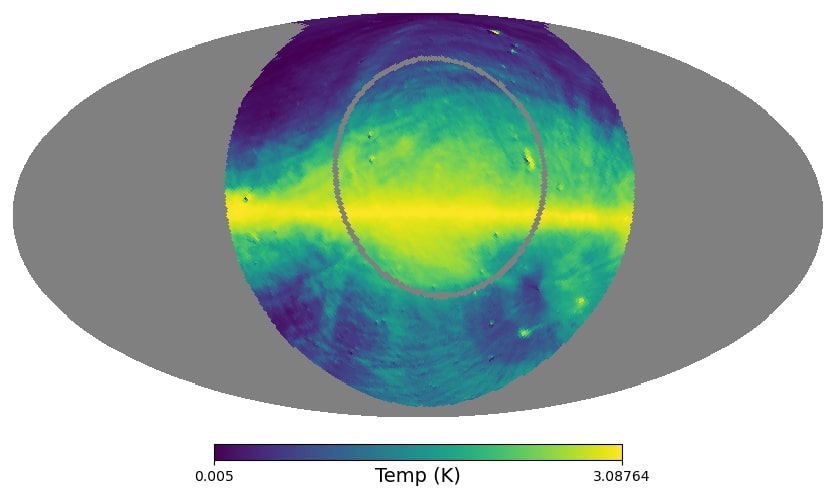}
    \caption{The beam foot print (at the full width half maximum point) on the sky shown as an encompassing grey `circle' at 2 GHz and 4 GHz, when the beam is wobbled 15$^{\circ}$ over the frequency range.}
    \label{fig:Beam_wobble_half_power}
\end{figure}

\begin{figure}[h]
    \centering
    \includegraphics[width = 0.9\linewidth]{Beam_wobble_-10deg_with_arrow.jpg}
    \includegraphics[width = 0.9\linewidth]{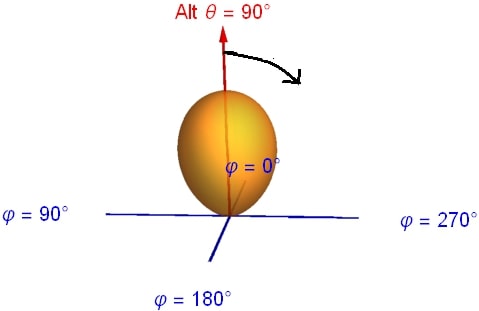}
    \includegraphics[width = 0.9\linewidth]{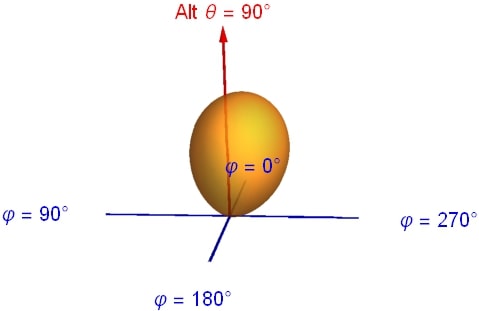}
    \caption{2-directional beam wobble with a total wobble angle of 10 $^{\circ}$ shown over spot frequencies of 2GHz, 3GHz, and 4GHz}
    \label{fig:2D_wobble_beam}
\end{figure}

Figure \ref{fig:Wobble_beam} shows the beam at 2 GHz, 3GHz and 4 GHz in the case of 1-D beam wobbling. This is also shown in the animation \href{https://tinyurl.com/46m25zwx}{https://tinyurl.com/46m25zwx} Mock sky spectra are then generated for varying amounts of wobble, i.e. for a range of $\alpha$. The effect of the beam wobble on the effective sky region observed is shown in Figure \ref{fig:Beam_wobble_half_power} and the corresponding animation is shown in the link at \href{https://tinyurl.com/2fd3479t}{https://tinyurl.com/2fd3479t}. It is clear that different parts of the sky are weighted differently depending on the frequency of observation and this results in spectral features in the mock spectrum and consequently changes the residual on fitting the same with an MS function. Mock spectra with perturbation are generated with the CRR signal and without to provide test residuals which are compared with the reference residual using the two detection metrics.
The treatment is similar for the 2-D beam wobble case, with the difference that the tilt direction reverses with frequency about the central frequency of 3 GHz, as shown in Figure \ref{fig:2D_wobble_beam}. The animated case of this 2-D wobble toy model is shown \href{https://tinyurl.com/ykfh6tb3}{https://tinyurl.com/ykfh6tb3}. 

\begin{figure}
    \centering
    \includegraphics[width = 0.9\linewidth]{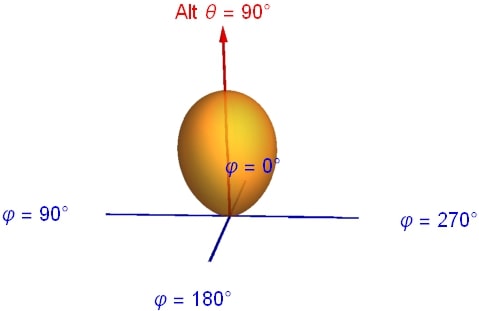} 
    \caption{Ideal $sin^2(\theta)$ beam}
    \label{fig:enter-label}
\end{figure}

\subsection{Antenna beam `stretching'}
The ideal $sin^2(\theta)$ beam has a full width at half maximum (FWHM) of 90 deg. In this perturbation the FWHM changes, once again weighting the foreground (pixels in the simulation) differently with frequency, always retaining the direction of the maximum towards zenith (90 $^{\circ}$ altitude). This can result in the beam looking squeezed or stretched. The stretching perturbation is parameterised by $\beta$, wherein the beam has a $sin^2(\theta)$ value or FHWM of 90$^{\circ}$ at the centre frequency of 3 GHz. For a particular value of $\beta$, we perturb the beam such that the FWHM of the beam increases smoothly for the frequencies from 2-4GHz. Thus, the beam looks squeezed at frequencies below 3 GHz, and stretched above.  For a particular value of maximum stretching, the beam weight across frequencies (freq) in the toy model is given by equation \ref{eq:stretch}. 

\begin{equation}\label{eq:stretch}
G(\theta, \phi, \nu) = \mathrm{sin}^2(\theta) + \beta(\nu_{GHz} - 3). \mathrm{sin}^2(2\theta)
\end{equation}

\begin{figure}[h]
    \centering
    \includegraphics[width = 0.9\linewidth]{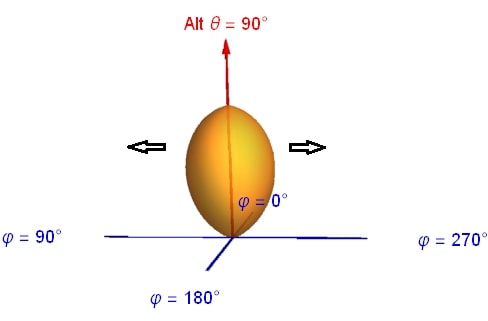}
    \includegraphics[width = 0.9\linewidth]{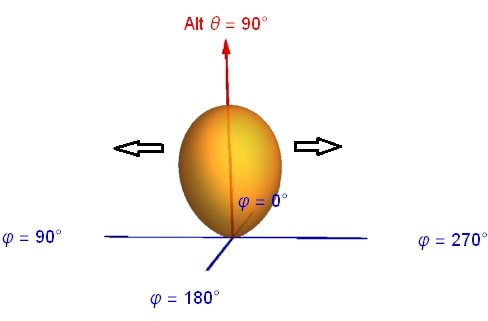}
    \includegraphics[width = 0.9\linewidth]{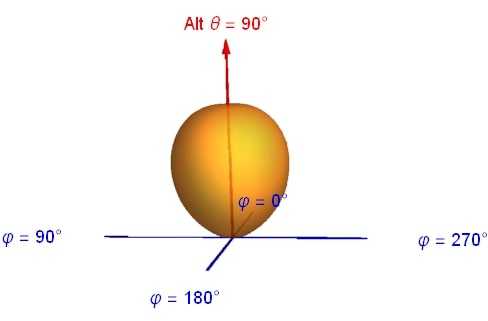}    
    \caption{Stretched beam from the toy model, 
 $\beta = 0.1$ $(\Delta\text{FWHM = 22.2}^{\circ})$}
    \label{fig:stretched_beam}
\end{figure}

\begin{figure}[h]
    \centering
    \includegraphics[width = \linewidth]{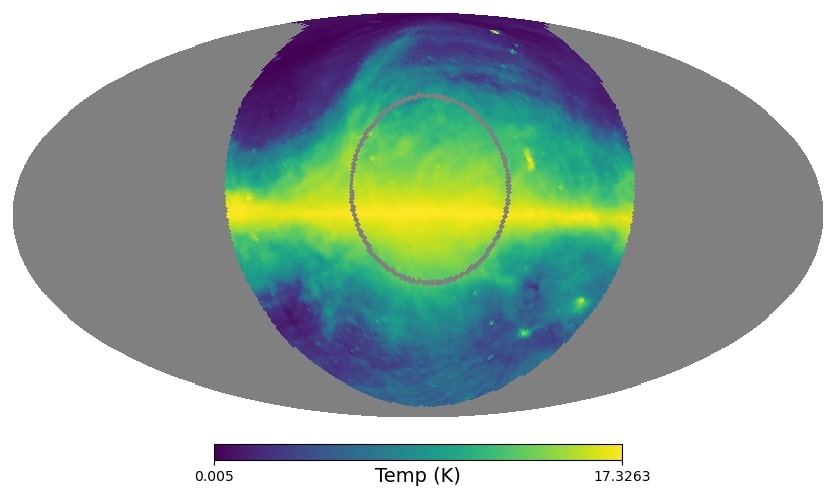}
    \includegraphics[width = \linewidth]{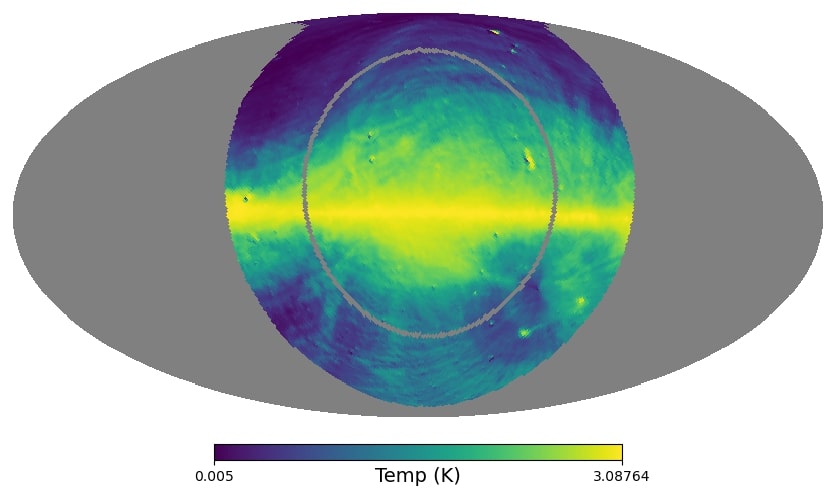}
    \caption{Half power beam width region comparison on the sky shown as an encompassing grey `circle' at frequencies of 2 GHz and 4 GHz, when the beam is stretched with $\beta$ = 0.2 $(\Delta\text{FWHM = 41.1}^{\circ})$ in the frequency range.}
    \label{fig: Beam_stretch_half_power}
\end{figure}

The perturbed beam at spot frequencies is shown in Fig \ref{fig:stretched_beam} and the resultant field of view from the changing beam footprint on the sky is in Figure \ref{fig: Beam_stretch_half_power}. An animation of beam stretching with $\beta = 0.2$ is at \href{https://tinyurl.com/5c84fnyx}{https://tinyurl.com/5c84fnyx} with the corresponding footprint on the sky at \href{https://tinyurl.com/39rduvvs}{https://tinyurl.com/39rduvvs}.

Varying the $\beta$ parameter to a maximum value of 0.25, we perturb the beam to have a maximum $\Delta$FWHM of 48$^{\circ}$ corresponding to an FWHM of $\sim 65^{\circ}$ at 2 GHz and $115^{\circ}$ at 4 GHz. Figure \ref{fig:stretched_beam} shows the beam at 2 GHz, 3GHz and 4 GHz with an FWHM variation of ${22.2}^{\circ}$ between them corresponding to  $\beta=0.1$. The field of view of the beam for $\beta=0.2$  $(\Delta\text{FWHM = 41.1}^{\circ})$ at both frequencies is shown in Figure \ref{fig: Beam_stretch_half_power}.

\section{Results}\label{sec:results}
With the beam perturbed as described in Section \ref{sec:perturbation}, mock sky spectra are generated for varying observation locations and LSTs. In each case one set of spectra are generated with the CRR signal included in them, and another without. The mock spectra so generated are then fit with a Maximally Smooth function of order 10 (or any arbitrarily large order) to generate residuals which serve as the test cases for further analysis. The two locations of choice are Bangalore and Murchison to provide varying sky coverage from the Northern and Southern hemisphere. Three representative LSTs chosen are at 17h (corresponding to Galactic centre overhead at Murchison), 9 hours (Galactic anticentre), and 13 h (somewhere in between the two). The variation in observing location and LST can help discern the effect of beam chromaticity on the observed sky spectrum and consequently interpretation of the detection metric.

\subsection{1-direction beam wobble}
\begin{figure}[h]
    \centering
    \includegraphics[width = \linewidth]{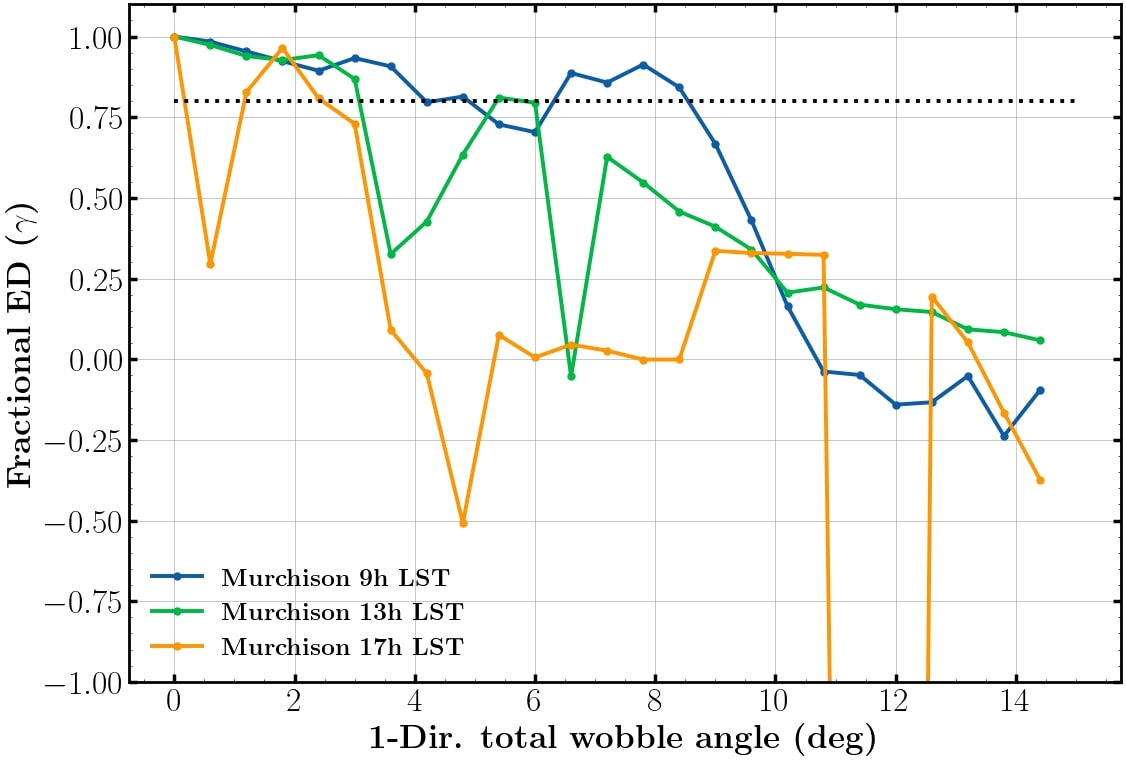}
    \includegraphics[width = \linewidth]{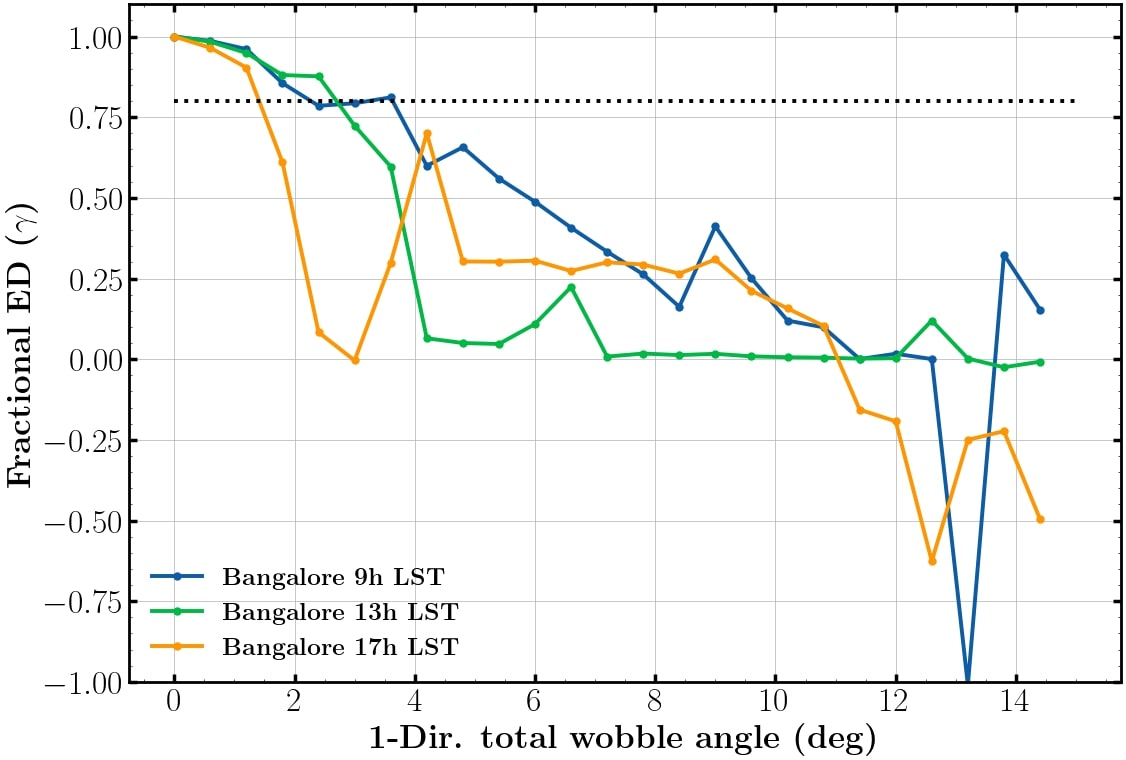}
    \caption{Fractional Euclidean distance ($\gamma$) comparison for different LSTs at Murchison and Bangalore for 1-D beam wobbling.}
    \label{fig:ED_graph_wobble}
\end{figure}

\begin{figure}[h]
    \centering
    \includegraphics[width = \linewidth]{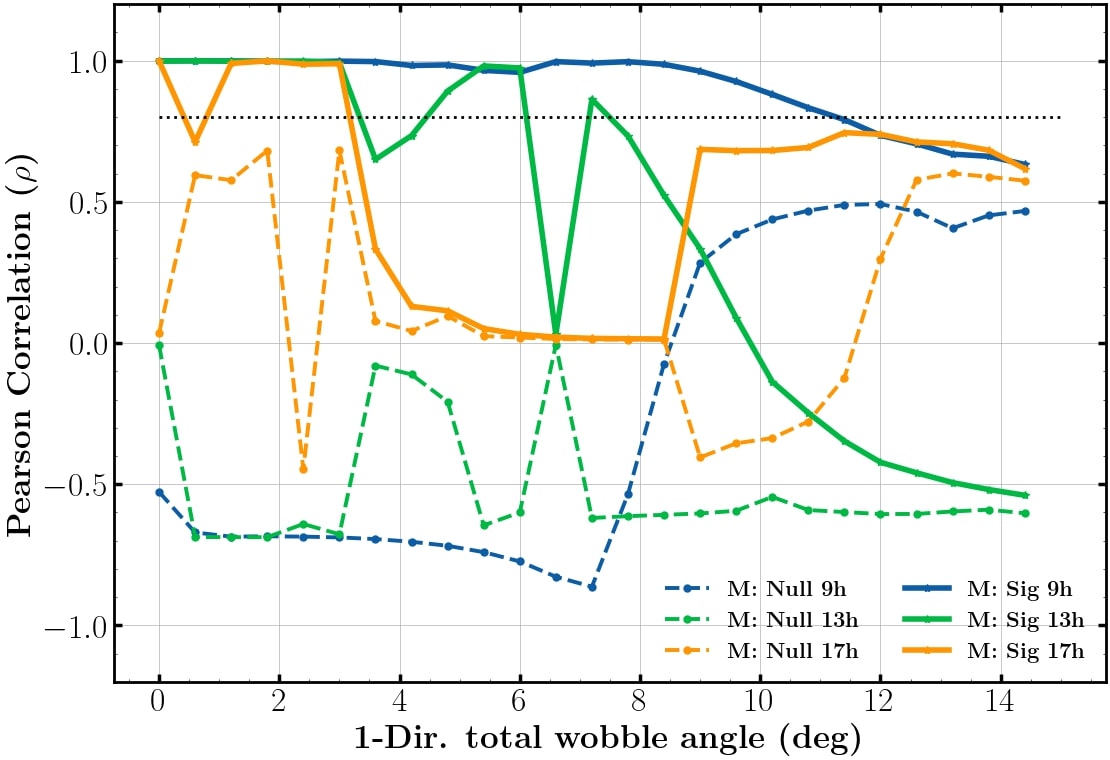}
    \includegraphics[width = \linewidth]{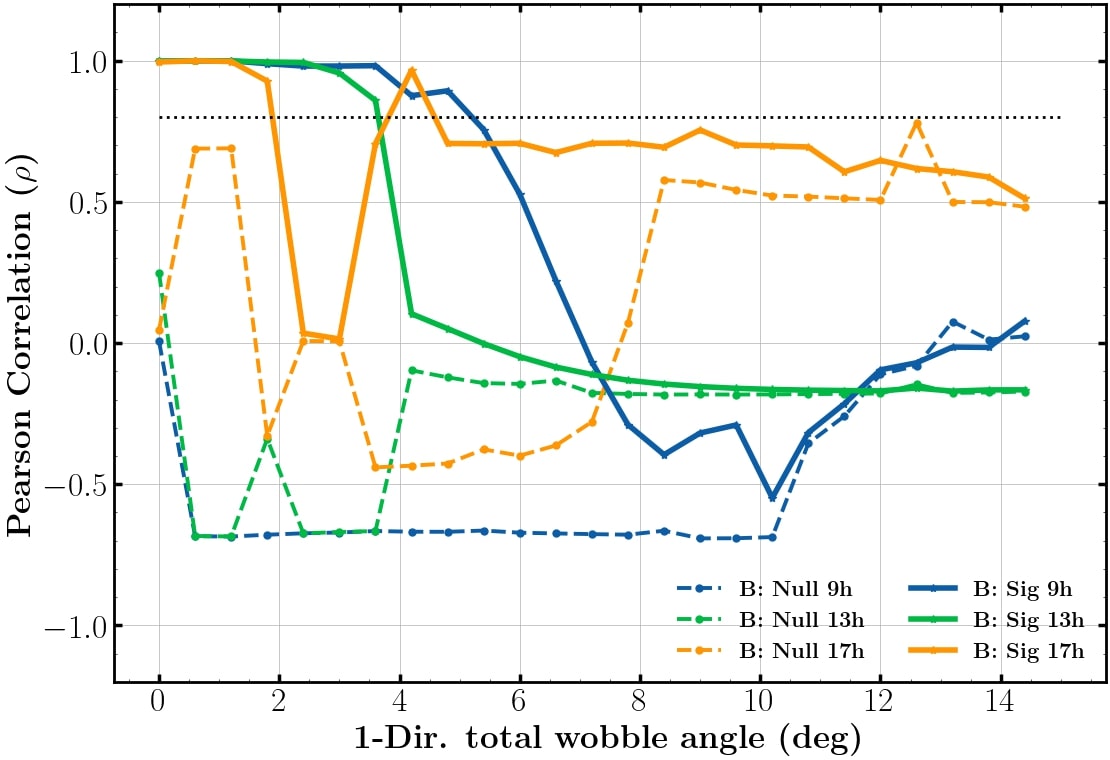}

    \caption{1-D wobble: Pearson Correlation coefficient ($\varrho$) comparison between recombination signal injected vs null case, at Murchison and Bangalore. Black dotted line marks the $\varrho = 0.8$}
    \label{fig:1d_wobble_PCC__2_locs}
\end{figure}

We examine the performance of the two metrics - fractional Euclidean distance and Pearson Correlation Coefficient - for the case when the beam wobbles in 1-Direction, i.e the beam maximum moves in one direction over the 2-4 GHz frequency range. The fractional Euclidean distance ($\gamma$) is in Figure \ref{fig:ED_graph_wobble}. At the lowest level of perturbation, $\alpha=0$ the the total wobble angle = 0, and this is identical to the ideal case, without any chromaticity. As expected for this case $\gamma=1$. As the perturbation is increased, $\gamma$ is $\lesssim 1$ for small perturbations and $\gamma <<1$ for large perturbation. This is the expected trend, as the residual departs further in shape and importantly in amplitude from the reference case as the beam introduces more structure in the mocks spectrum. In the limiting case, the residual with the CRR signal and without in the test case are identical to one another and very different from the reference case, resulting in $\gamma=0$. Placing an threshold of $\gamma=0.8$ to claim a detection, we make the following observations. With a non-ideal beam exhibiting 1-directional wobbling behaviour with frequency, the range of beam non-ideality that is tolerable for a signal detection is finite, where in one can distinguish between signal presence and absence. The tolerance is a function of observing location and LST, with it being largest (4.8$^{\circ}$) at LST 9h in Murchison, looking straight away from the Galactic centre. The tolerance is lowest (0.6$^{\circ}$) at LST 17h in Murchison when the Galactic centre is overhead and any change in the beam behaviour translates more significantly into features in the spectrum.  The corresponding Pearson Correlation Coefficient is shown in figure \ref{fig:1d_wobble_PCC__2_locs}. A few observations can be made right away. For all locations and LSTs, at zero wobble, $\varrho=1$ for the test cases with the CRR signal present and $\varrho=0$ when it is absent. Again using 0.8 as a threshold to distinguish between the two cases, the tolerance in the acceptable beam wobble to make a distinction between the signal present and absent in the test cases is larger than when using $\gamma$. This is as expected, given $\varrho$ is a more relaxed metric of detection, as it is a more qualitative comparison of the test and reference cases. The trends in the location and LST are similar to $\gamma$ and in most cases the tolerance is finite, and non-zero. The underlying the effect of the change in the beam direction on the sky has on signal detection is evidenced by the richness in the $\varrho$ curves, providing insight into choosing optimal observing locations and LSTs.

\begin{figure}[h]
    \centering
    \includegraphics[width = \linewidth]{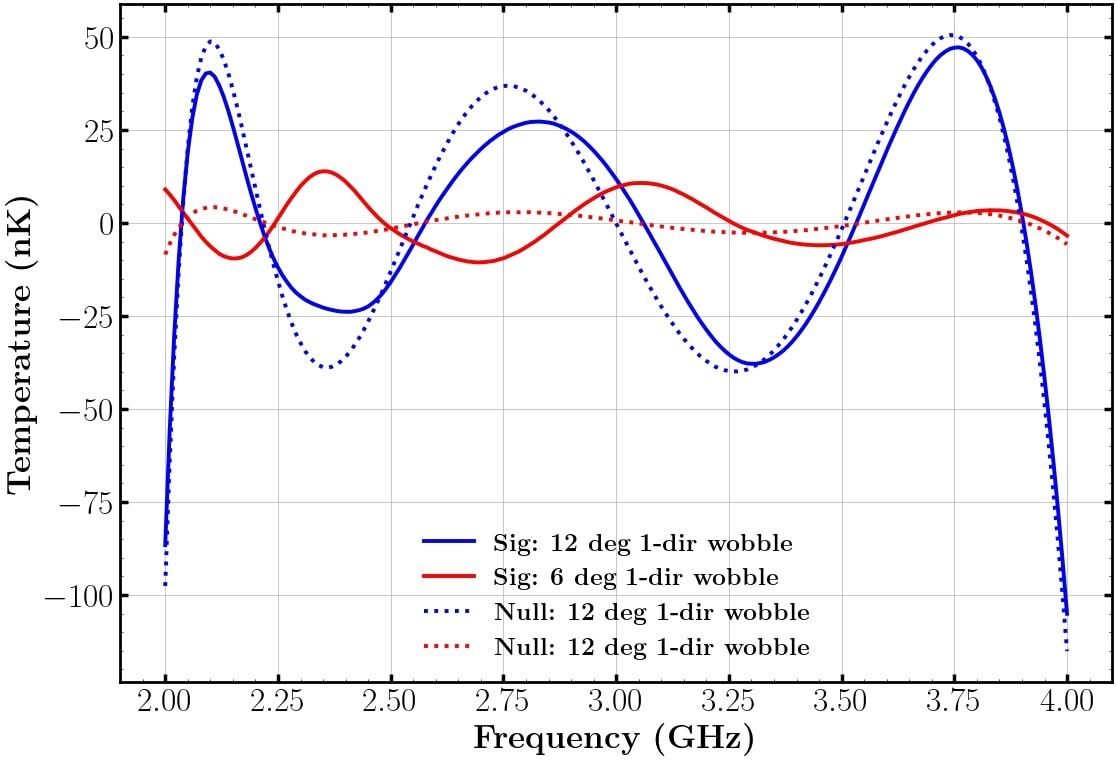}
    \caption{The effect of 1-D wobble beam perturbation at Murchison, 13hrs LST. Increasing the perturbation increases the residuals, suggesting leakage of spatial foreground features into spectral structures.}
    \label{fig:ED_graph_wobble_various_cases}
\end{figure}

\subsection{2-direction beam wobble}
\begin{figure}[h]
    \centering
    \includegraphics[width = \linewidth]{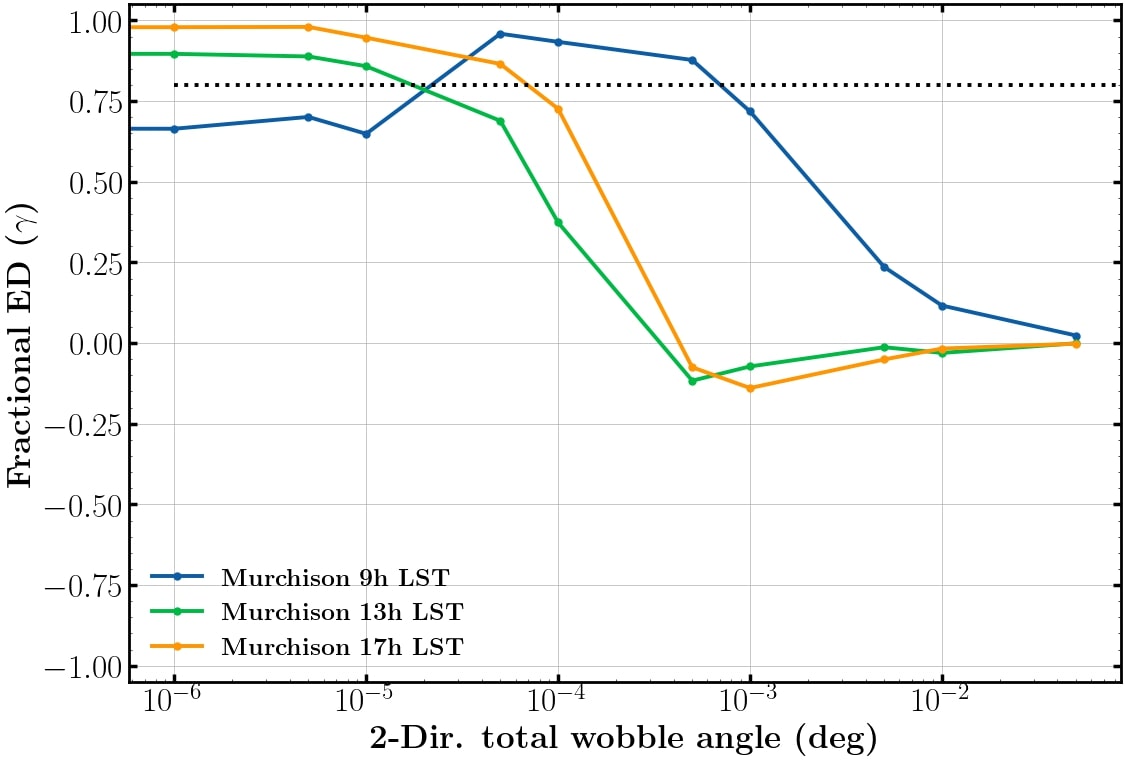}
    \includegraphics[width = \linewidth]{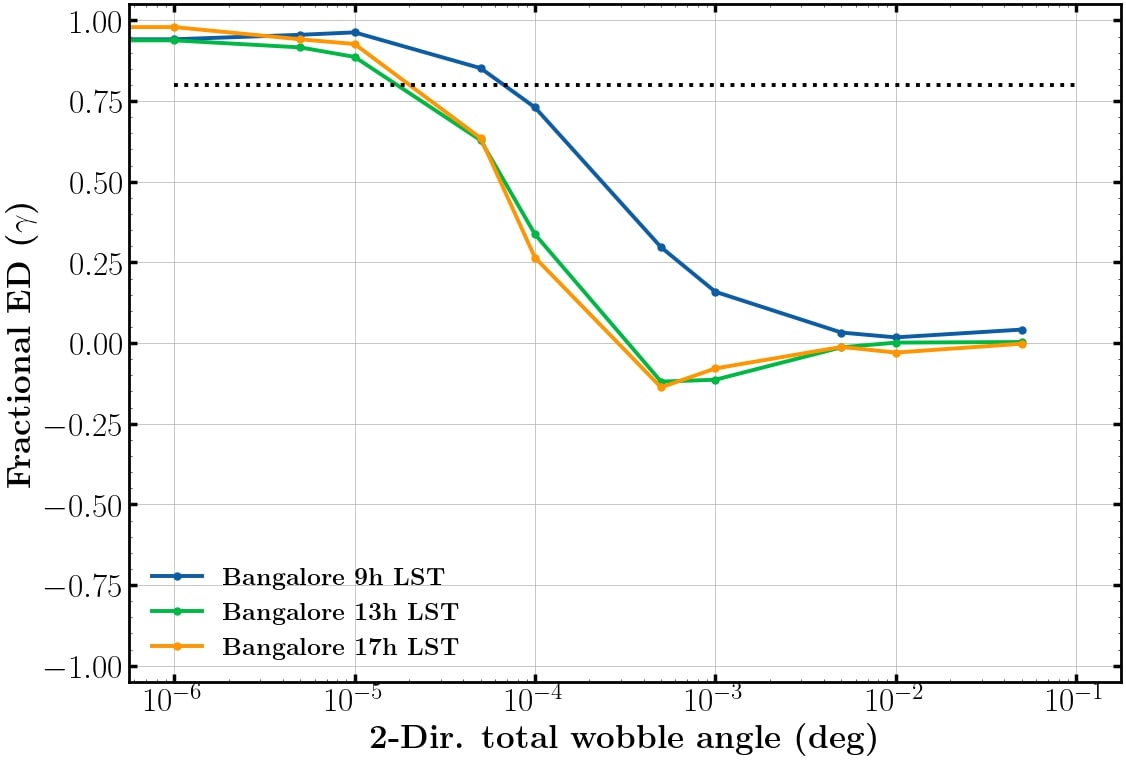}
    \caption{Fractional Euclidean distance ($\gamma$) for different LSTs at Murchison and Bangalore for 2-D beam wobbling. Note that the 2-D total wobble angle is in the log scale.}
    \label{fig:ED_graph_2dwobble}
\end{figure}

\begin{figure}[h]
    \centering
    \includegraphics[width = \linewidth]{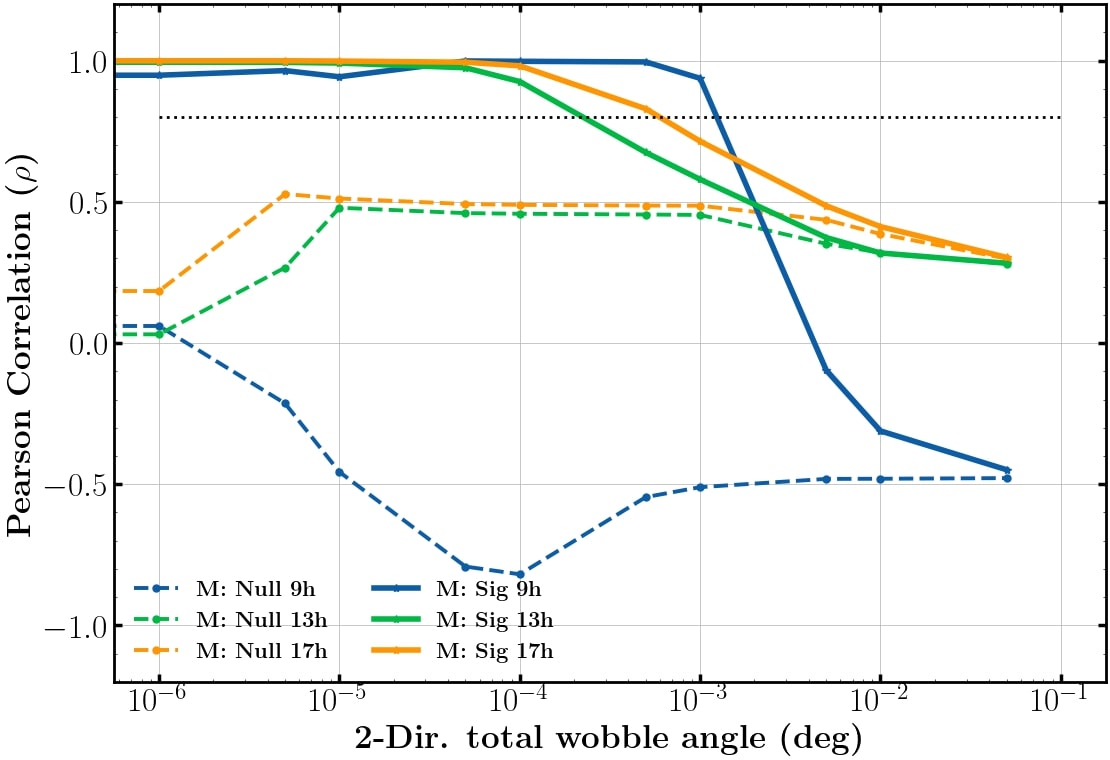}
    \includegraphics[width = \linewidth]{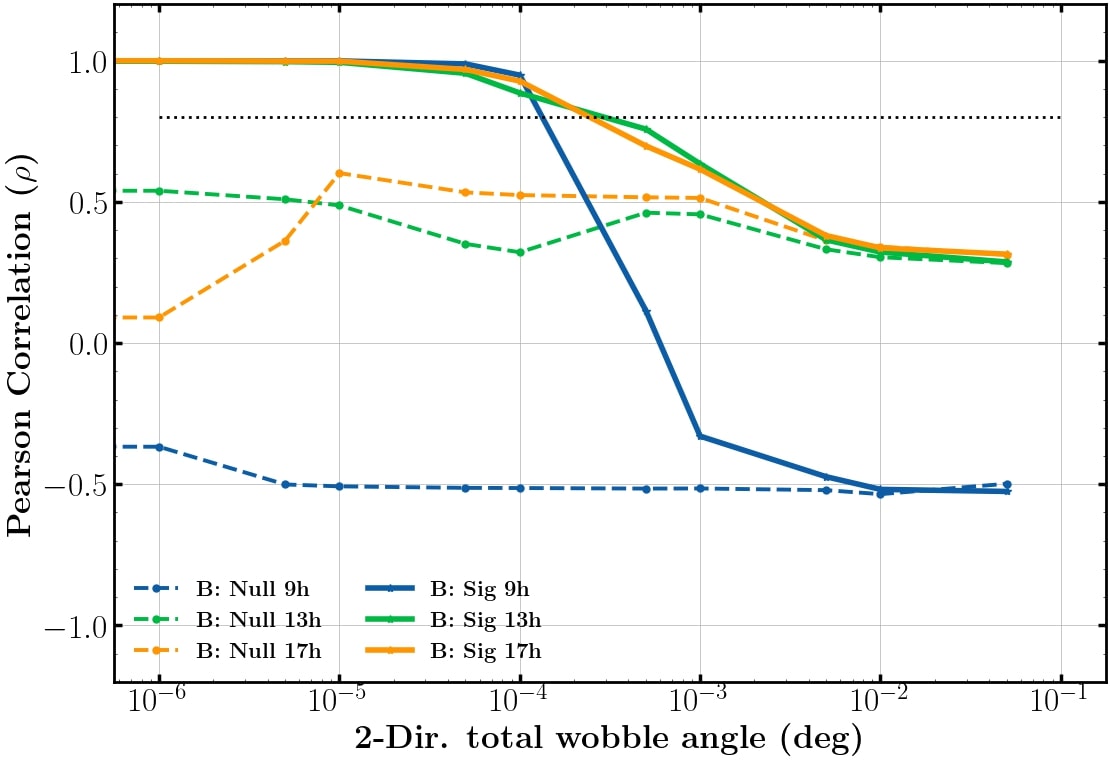}

    \caption{2-D wobble: Pearson Correlation coefficient ($\varrho$) comparison between recombination signal injected vs null case, at Murchison and Bangalore. Note that the x-axis is log-scaled. Black dotted line marks the $\varrho = 0.8$}
    \label{fig:2d_wobble_PCC_2_locs}
\end{figure}

The $\gamma$ and $\varrho$ when the beam is perturbed to have a shift in the direction of peak wobble about the centre frequency, i.e. 2-D wobble, are shown in Figures \ref{fig:ED_graph_2dwobble} and \ref{fig:2d_wobble_PCC_2_locs} respectively. Using the same threshold of signal detection as in the 1-D wobble case, we make the following observations. The tolerance on 2-D beam shift is much tighter than the 1-D case, which can be attributed to the intrinsic non-linearity in the beam's frequency dependence. This is attributed to the discontinuity in the beam behaviour at 3 GHz. If one were to necessarily use the full band, this non-smooth feature at the band center will always result in unphysical spike like features in the residuals. Thus, tolerances in simulation, though finite, can be considered negligible. If the frequency of the beam direction change is closer to a band edge, i.e. either closer to 2 GHz or 4 GHz, the case would be closer to a 1-D wobble over a significant portion of the band, and only the contiguous stretch of frequency where the beam direction is maintained can be processed. This can help informing antenna design, either in avoiding peak direction shift entirely, or maintain the direction of beam shift change close to the band edge. Other treatments such as flagging the turnover frequency may also be employed. In either case, the best way to utilize data from an experiment where the beam changes direction is by dropping channels, and more the number of such direction changes, the larger the fraction of channel loss would be, resulting in lower confidence in detection. 

Investigating the different types of beam wobble (1-D vs 2-D) suggests that it is more suitable to have an antenna with ``smooth" changes in its beam systematics across frequencies. The abrupt change in the wobble direction leads to a noticeable departure from overall spectrum smoothness and is best avoided. 

\subsection{Beam stretching}
\begin{figure}[htb]
    \centering
    \includegraphics[width = \linewidth]{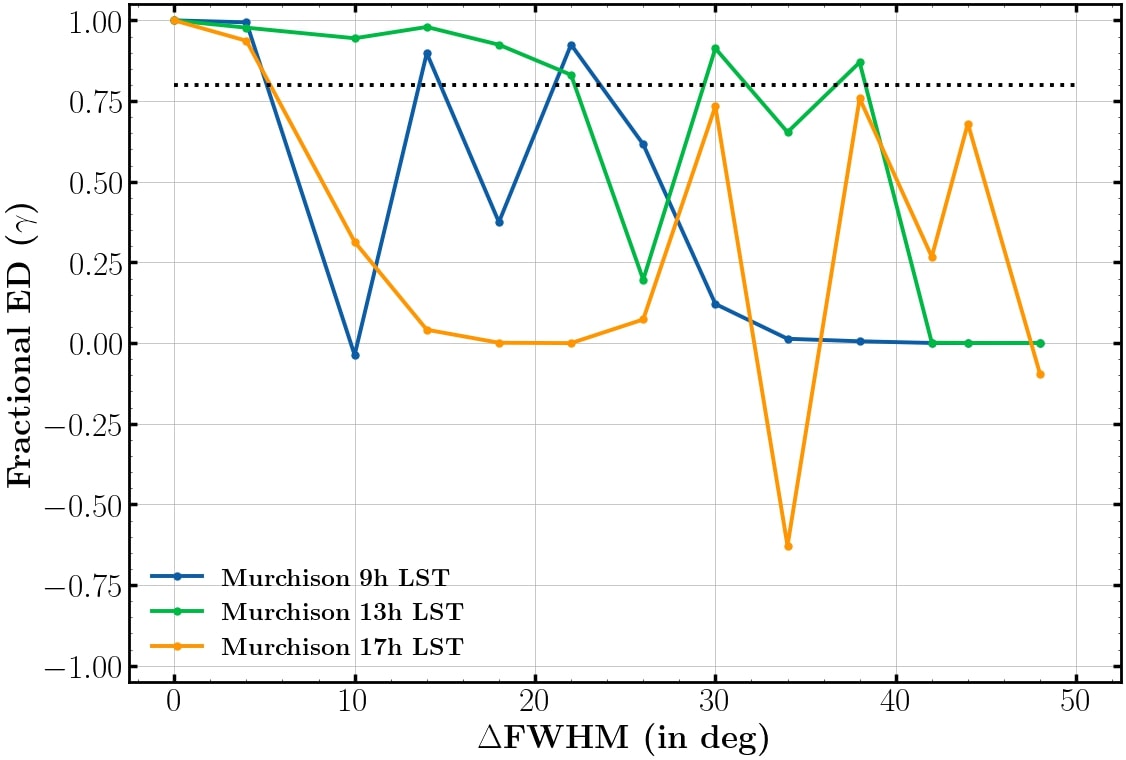}
    \includegraphics[width = \linewidth]{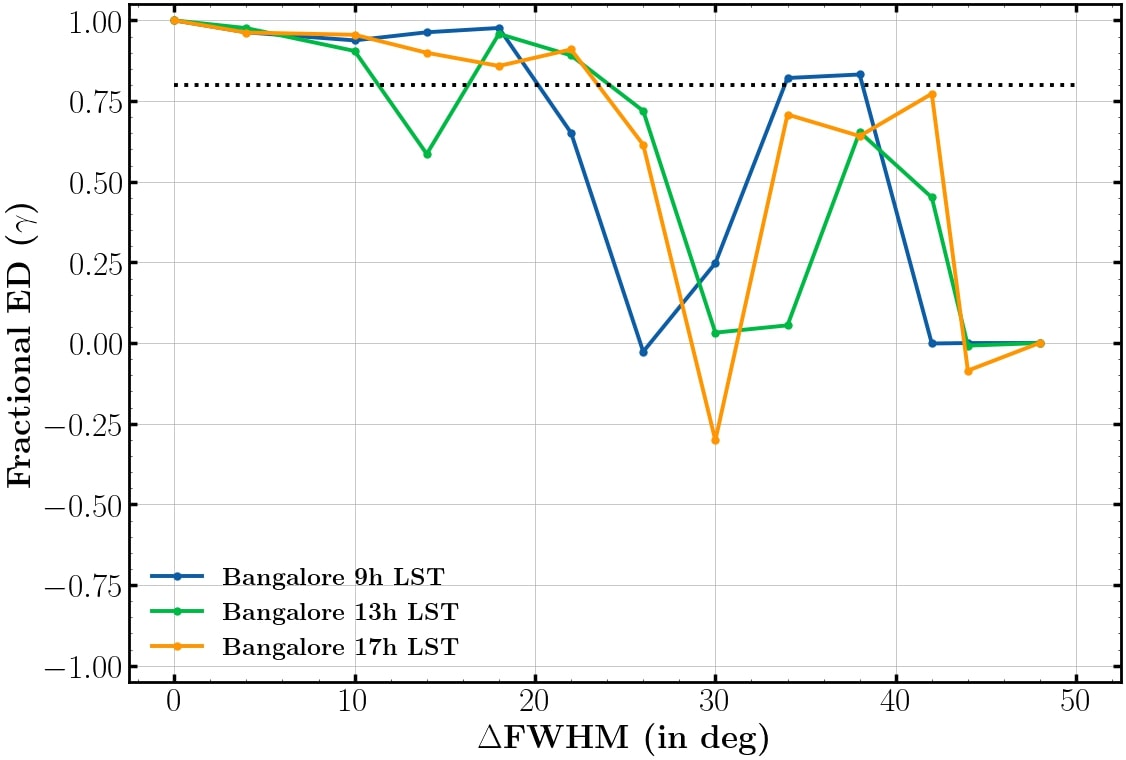}
    \caption{Fractional Euclidean distance ($\gamma$) comparison for different LSTs at Murchison and Bangalore for beam stretching.}
    \label{fig:ED_graph_stretch}
\end{figure}

\begin{figure}[ht]
    \centering
    \includegraphics[width = \linewidth]{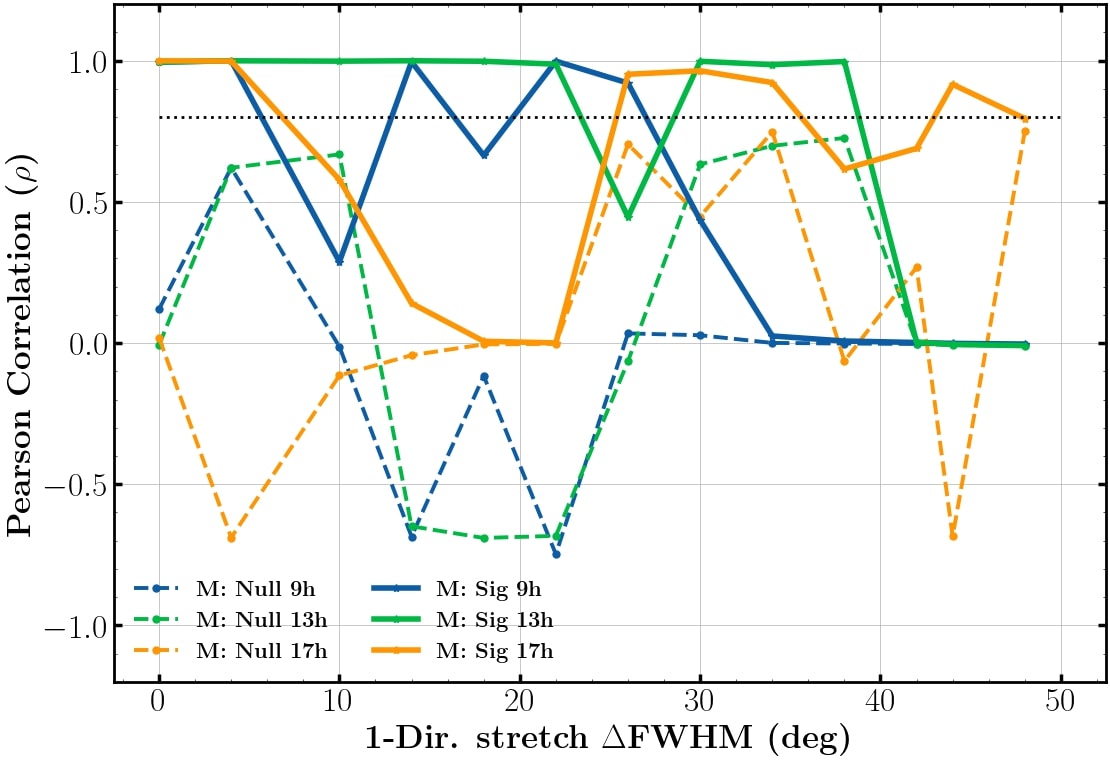}
    \includegraphics[width = \linewidth]{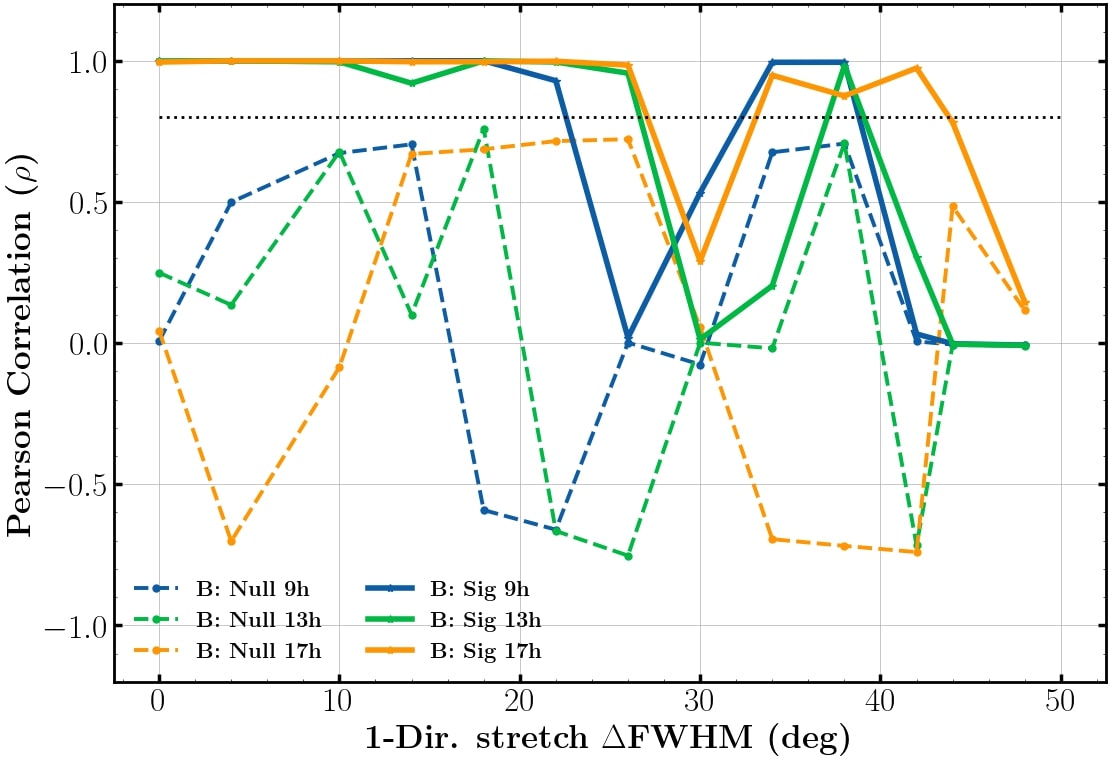}

    \caption{Beam stretching - Pearson Correlation coefficient ($\varrho$) comparison between recombination signal injected vs null case, at Murchison and Bangalore.}
    \label{fig:1d_stretch_PCC__2_locs}
\end{figure}

The metrics of signal detection for test cases when the beam is stretched with increasing frequency are presented in Figures \ref{fig:ED_graph_stretch} and \ref{fig:1d_stretch_PCC__2_locs}. Some observations follow. Once again, at all locations when the stretching parameter $\beta=0$, the test case with the signal is identical to the reference case resulting in $\gamma=1$ and $\varrho=1$, and the test case without signal is noise, reflected by $\varrho=0$. The tolerance on beam stretching at all locations and LSTs is finite and significantly larger than the best cases for 1-D beam stretch. This is expected, as the the direction of beam maximum is always fixed towards zenith, i.e. towards the same pixel. The effect of beam stretching only affects the pixels that are naturally weighted lower resulting in the spectral structure they introduce to be down weighted. Whereas the tolerance in general is larger for Bangalore compared to Murchison which sees the Galactic centre overhead, it is interesting to note that the LST of 13 hours offers the most tolerance as the pixels that do enter the beam because of stretching are more likely to be cooler and thus have a lower effect on the final spectrum when observed through the non-ideal beam.

A summary of the tolerances in the different types of beam perturbation for a metric threshold of 0.8 is in Table \ref{tab: tolerances}. The tolerances for the 1D and 2D wobble represent the total wobble angle permitted until the threshold drops to or below 0.8 at the first instance, and the tolerance for the stretch represents the change in the FWHM ($\Delta FWHM$) permitted until the threshold drops to or below 0.8 at the first instance. The locations are indicated as M for Murchison, and B for Bangalore.

\begin{table}[hb]\label{tab: tolerances}
\tabularfont
\setlength\tabcolsep{1pt} 
\caption{Tolerances in beam perturbations}\label{tableExample} 
\begin{tabular}{lcccccc}
\topline
type&M9($^{\circ}$)&M13($^{\circ}$)&M17($^{\circ}$)&B9($^{\circ}$)&B13($^{\circ}$)&B17($^{\circ}$)\\\midline
1D-$\gamma$&4.8&2.8&0.6&3.6&2.4&1.2\\
1D-$\varrho$&11.4&2.8&0&4.8&3.6&1.8\\
\hline
2D-$\gamma$&0&0.5$\times10^{-4}$&$10^{-4}$&$10^{-4}$&$0.5\times10^{-4}$&$0.5\times10^{-4}$\\
2D-$\varrho$&$10^{-3}$&$0.5\times10^{-3}$&$10^{-3}$&$10^{-4}$&$0.5\times10^{-3}$&$0.5\times10^{-3}$\\
\hline
stretch-$\gamma$&4&18&4&18&10&22\\
stretch-$\varrho$&4&22&4&22&26&26\\
\hline
\end{tabular}
\end{table}

\section{Conclusion}\label{sec:conclusion}
We have demonstrated that using standard measures of similarity as signal detection metrics, it is feasible to distinguish between the presence and absence of CRR lines despite adopting a non-ideal beam for spectrum simulation. The fractional Euclidean distance serves as a conservative metric and thus a stronger measure of signal detection permitting tighter tolerances in beam perturbations. Using the the Pearson correlation coefficient allows a larger range in the beam perturbations while discerning between signal presence and absence, which may serve as a first step towards detection while improving antenna properties. While the beam perturbations discussed herein are simplistic, it is a first of its kind approach, moving a step ahead towards a practical treatment towards CRR signal detection, a step forward from the ideal instruments in all simulations thus far. Future work will consider more complex chromatic features in the antenna and full system. The results from this study provide confidence that there are certain types of beam chromaticity (such as 1-D wobble, and stretching) that provide a larger margin of tolerance for signal detection, whereas other kinds of chromaticity (2-D wobble) are more detrimental to signal detection by way of channel loss. The choice of observing location and the LST play an important role in the way beam chromaticity affects prospects of signal detection. The signal detection metrics themselves rely on the well predicted theoretical expectation of the CRR signal, which is a powerful tool that can be harnessed for experiment design. Any detection is claimed by being able to distinguish from a null hypothesis scenario, which can be realized in practise using appropriate calibration strategies, such as the use of an external blackbody or external `mock sky' source. All the above can help inform antenna modeling, observing and calibration strategy, and experiment design, bringing us one step closer to detecting the CRR line, an inevitable prediction of standard cosmology.

\section*{Acknowledgements}
The authors thank Dr. Anjan Sarkar for discussions and valuable inputs.


\bibliography{ref}





\end{document}